\documentclass[aps,prb,preprint]{revtex4-1}
\usepackage{graphicx}
\usepackage{bm}
\usepackage{epstopdf}
\usepackage{color}

\begin{document}

\title{The structure of Lennard-Jones nanowires encapsulated by carbon nanotubes}

\author{W. Q. Wu, Q. H. Yuan and D. Y. Sun}
\address{Department of Physics, East China Normal University,
Shanghai 200062, China}

\date{\today}

\begin{abstract}
Molecular dynamics simulations have been used to investigate the structures
 of Lennard-Jones (LJ) nanowires (NWs) encapsulated by carbon nanotubes (CNTs). 
 We found that, when the radius of CNTs is small, the structures of NWs are quite simple, 
 {\it i.e,} only multishell NWs are formed. For CNTs with larger radius, the structure of
  NWs becomes much richer. In addition to crystal and multishell NWs, three new kinds of NWs 
  are found,  {\it i.e,} a hybrid NW with a crystal core coated by a few shells, a crystal NW with 
  droplet-like pits on the side of NW, and a multishell NW with droplet-like pits on the side of NW. 
   The 'phase' diagram, which describes the structure change with the number of LJ atoms and the
    interaction between NWs and CNTs, is also obtained.
\end{abstract}


\maketitle

 \section{INTRODUCTION}

CNTs filled with various materials are of great interests  in nanoscience and nanotechnology.
\cite{APcata,APcata2,APcata3,APcata4,APele,APele2,APdata,APsen,APosc,APosc2}  As a special 
one-dimensional system,  the material filled CNT usually has many exotic physical and chemical 
properties. For example, the NWs encapsulated by CNTs have extremely high chemical and 
mechanical stability, which ensure that the NWs inside CNT can be used in various environments.
 The superior properties endowed NWs with promising applications in  heterogeneous catalysis,
 \cite{APcata,APcata2,APcata3,APcata4} electromagnetic wave absorption,\cite{APele,APele2}
  high density magnetic data storage devices,\cite{APdata} sensors for capacitive humidity,
  \cite{APsen} and oscillators with high frequency,\cite{APosc,APosc2} {\it etc}.

Previous studies have shown that the material encapsulated  by CNTs could have much different 
structures from that of its freestanding counterpart
\cite{FSzr,FSau,FSau2,FSpd,MSau,MSau2,MSna,MScs,MSfe,MSau3,EPpar,EPau}. The reason is that the elastic 
energy caused by mismatch between intrinsic length scale of solid (bond length, lattice constants, 
{\it etc}) and the dimension of confinement space has important effect on the structures or 
morphologies of NWs. Physically, the elastic energy can be mainly determined by three important 
factors, the number of filled atoms, the radius of CNTs, as well as the interaction between NWs 
and CNTs.

Many studies have shown that, the structure of NWs encapsulated  by CNTs (or generally a tube) could be effected 
by many factors, such as the radius of CNTs, the number of filled atoms or their interactions with CNTs. 
\cite{PTcu,PTau,PTtm,PTga,PTcu2,Peeters1,Peeters2}  For examples, the effect of CNT radius on the structure of encapsulated
 NWs has been investigated by  Choi {\it et al}, they found that the stable structure of copper 
 confined inside CNT is multishell packing when the diameter of CNTs is small, but a 
 face-centered-cubic (fcc) structure as the diameter of CNT increases.\cite{PTcu} 
 A similar study has also shown that, the number of filled Au atoms plays a dominant role
 in the final structure of gold  encapsulated by CNTs.\cite{PTau} A multishell NW will be 
 formed as the number of filled Au atoms is large enough, otherwise a crystal NW will appear. 
 The effect of interaction strength between the transition metal atom and CNTs has been explored
  by using Monte Carlo simulations. The simulations found that the structure of transition 
  metals encapsulated by CNTs was strongly dependent on the interaction strength.\cite{PTtm} 
  However, the combination effect of all the three major factors on the structure of NWs 
  stays unclear.

In the present work, the structure of LJ NWs encapsulated by CNT is investigated by using 
molecular dynamics (MD) simulations. Three important factors,  the number of encapsulated LJ 
atoms, the interaction between LJ atoms and CNTs, as well as radius of CNTs,  are considered 
together to explore their effect on the structure of a NW encapsulated by CNTs.  Our simulation
 results show that, NWs encapsulated by CNTs with small radius  have simple structures, 
 while NWs encapsulated by CNTs with large radius have much richer structures. Three new 
 structures are identified for NWs encapsulated by large radius CNTs, and hence the phase 
 diagram describing the structure change with the number of LJ atoms and the interaction
  between NWs and CNTs  was improved. Our results also shed lights on the structure of a solid
   confined inside a nano pore.

\section{Computational Details}

Two single walled CNTs, with an index of (15,15) or (30,30) respectively, are studied. The length of CNTs is around 10 nm, corresponding to 2460 and 4920 carbon atoms in (15,15) and (30,30) CNT respectively. The periodic boundary condition along the axial direction of CNTs is used and free boundary conditions are applied in the radial directions. This implies that the system is infinite along the axial direction.  Similar to most previous studies,\cite{MSna,MScs,PTcu,PTtm} CNTs were treated as a rigid tube.  All LJ atoms are confined inside CNTs, no atom is allowed at the outside of CNTs. The  number of LJ atoms ($N_{LJ}$) is varied from 1500 to 6000 in (30,30) CNT and from 200 to 1200 in (15,15) CNT.

The Adaptive Intermolecular Reactive Empirical Bond Order Potential with intermolecular interactions,\cite{CH,CH2} which is widely used in previous studies, is adopted to model carbon atoms in CNTs. Since we are interested in general results for NWs encapsulated by CNTs, the truncated  LJ potential\cite{LJ} is adopted to describe the interatomic interaction between LJ atoms, as well as the LJ atom-CNT interactions, which has been used for similar systems previously.\cite{Cu} One potential parameter between LJ atoms and carbons ($\sigma_{sc}=3.805\AA$, determined according to the Lorentz-Berthelot mixing rules), characterizing the interaction length scale between carbon and LJ atoms, is fixed through the whole paper.  Another mixed potential parameter ($\epsilon_{sc}$), describing the interaction strength between carbon and LJ atoms, is changeable to tune the interaction strength. In our present work, a weaker interaction ($\epsilon_{sc}$=0.015eV), a normal interaction ($\epsilon_{sc}$=0.02578eV, determined according to the Lorentz-Berthelot mixing rules) and  a stronger interaction ($\epsilon_{sc}=$0.035eV) are chosen to investigate the interaction effect on the structure of encapsulated LJ NWs.

All MD simulations made use of the LAMMPS (large-scale atomic/molecular massively parallel simulator) code with time step of 1fs\cite{Lammps}. The temperature is controlled by Nose-Hoover thermostat.\cite{NH} The procedure to get the stable structure at 0K is  based on the simulated annealing technique. Initially the system is equilibrated at 1500K, in which the  encapsulated atoms are in a well-defined liquid state. Then the system is cooled down to 0K with very low cooling rate. This process are repeated for a few times. Finally the stable structure is obtained among these structures. To characterize the structure, the pair correlation function (PCF) and angle correlation function (ACF) are calculated. These quantities are calculated in the usual way.

\section{RESULTS AND DISCUSSION}

 We have firstly studied LJ NWs encapsulated by (15,15) CNT. The results show that,   for all the studied cases here, NWs  take multishell structure. The multishell NWs have three coaxial cylindrical shells and a single atoms chain at the center. Each shell has a fcc (111) structure. Kondo and Takayanagi (KT)\cite{KT} have developed the so-called KT notation  $n-n^{'}-n^{''}-n^{'''}$ to describe a multishell NW consisting of coaxial shells with $n,n^{'},n^{''},n^{'''}$ atom rows.  Most of the multishell share the same KT index of 16-11-6-1, while a few others have a 17-12-6-1 KT index.

For LJ NWs encapsulated by (30,30) CNT, the structure of NWs become much richer, {\it i.e,} five structural motifs are found, in which three of them are not reported previously.  The first one with finite length, labeled as crystal NW, is shown in Fig. 1(a). Here the finite length refers the length of NWs less than that of CNTs, similarly hereinafter.  The crystal NW can be formed at weaker interaction ($\epsilon_{sc}=0.015eV$) with most filling number of LJ atoms or normal interaction (($\epsilon_{sc}=0.02578eV$) with relatively small $N_{LJ}$.  At stronger interaction ($\epsilon_{sc}=0.035eV$),  the crystal NW will not exist anymore for all $N_{LJ}$ studied here.

From Fig. 1(a), one can see that,  the surface atoms in crystal NW exhibit crystalline feature, and the cross section takes fcc (111) surface. In fact, the atomic structure of crystal NWs is based on fcc arrangement, which can also be seen from ACF and PCF shown in Fig. 2. Main peaks in both ACF and PCF match the fcc crystal well. In ACF,  there are three main peaks centered at $60^{o}$, $90^{o}$ and $120^{o}$, corresponding to the bond angle in a fcc crystal. However there are two small peaks around $110^{o}$ and $146^{o}$. The $110^{o}$ is related to the angle in twin boundary lying on a (111) plane.  The $146^{o}$ exits in the  ACF of hexagonal close packed  crystals, implying the existence of stacking faults. The PCF also shows some small peaks beside the main peaks of fcc crystal, indicating the slight distortion from fcc structure.

The second  structural type is the so-called multishell NW with finite length shown in Fig. 1(b). At weaker interaction ($\epsilon_{sc}$=0.015eV), the multishell NW does not exist for all $N_{LJ}$ studied here. At  normal and larger interaction ($\epsilon_{sc}$=0.02578 and 0.035eV), the multishell NW will be formed for higher filling fraction (larger $N_{LJ}$). From Fig. 1(b), one can see that, the multishell NW is composed of eight coaxial cylindrical shells. The spreading sheet of each shell are shown in upper panel of Fig. 3. Each shell exhibits a fcc (111) like structure. From the spreading sheets, one can see that, the KT index of the multishell NW is 43-38-33-28-21-16-10-4. The multishell NWs have been theoretically predicted for many different metals\cite{MSau2,MSna,MScs} and experimentally observed for Au NWs\cite{EPau}.  For the ACF and PCF of multishell NWs, which are shown in  Fig. 2, the width of all the peaks are broadened, and the distribution becomes nonzero among main peaks, which implies the large distortion on the fcc-based structure. Specially, the second peak of both ACF and PCF almost disappears.

The third structural type is a new one with finite length, labeled as hybrid NW. From Fig. 1(c), one can see that, in hybrid NWs, the LJ atoms near CNT walls form several shells, while the inner LJ atoms show crystalline feature. The hybrid NW only exists at larger interaction  ($\epsilon_{sc}$=0.035eV) and lower filling fraction (smaller $N_{LJ}$).  Similar structure has been reported for free-standing NWs and considered to be crystal inner core coated with several shells.\cite{FSau2}  However, the hybrid NWs inside CNTs have not been reported previously.  Although the positions of main peaks in both ACF and PCF (Fig. 2) are similar to fcc crystal, each peak is much broadened, indicating increased distortion from fcc structure. The spreading sheets of five outmost shells in hybrid NW are shown in lower panel of Fig. 3. Each outer shells in the hybrid NW are composed of both square (fcc (100) surface) and hexagonal (fcc (111) surface) lattices.  It  should be pointed out that, the structure of these shells is quite different from multishell NWs discussed above, in which all the shells in multishell NW are only composed of hexagonal lattices.

The forth one, labeled as crystal-pit NW, has the same length as CNT, actually becomes infinite length under the periodic boundary condition. The top and side view shows crystalline character (see Fig. 1(d)),  but two droplet-shape pits are formed near the wall. From ACF and PCF shown in Fig. 2, we can see the clear character of fcc structure. The peaks around $110^{o}$ and $146^{o}$  indicate that this NW also involves stacking faults. In fact, this NW has the similar atomic structure to that of crystal NW, besides of two droplet-shape pits near the wall. The crystal-pit NW only exists at weaker interaction ($\epsilon_{sc}$=0.015eV) and the highest filling fraction (the largest $N_{LJ}$ 5981) studied in current paper.

The last structural type obtained in our study is a multishell structure with two droplet-shape pits near the wall, labeled as multishell-pit NW. Besides of the droplet-shape pits, this NW are similar to multishell NWs previously discussed, it shares the same KT index  as multishell NWs.  The ACF and PCF (Fig. 2) are also very similar to multishell NWs. This NW also has the same length as CNT, actually becomes infinite length under the periodic boundary condition. The multishell-pit NW only exists at normal interaction ($\epsilon_{sc}$=0.02578eV) and the highest filling fraction (the largest $N_{LJ}$ 5981) studied in current paper. To our best knowledge, both crystal-pit and multishell-pit NWs were not reported before.

The phase diagram for NWs in (30,30) CNT is summarized in Fig. 4.  The result shows the stable structure of LJ NWs encapsulated by CNTs highly depends on $N_{LJ}$ and  $\epsilon_{cs}$.  For weaker interaction ($\epsilon_{sc}=0.015eV$), the atoms inside CNT formed a crystal NW with fcc structure and associated with grain boundaries as $N_{LJ}$ is less than or equal 5343.
Further increasing $N_{LJ}$, the crystal-pits NWs is formed when $N_{LJ}$ equal 5981.  For the normal interaction ($\epsilon_{sc}$=0.02578eV),  as $N_{LJ}$ less than or equal 3471, the crystal NW is formed. Multishell NW will be presented when $N_{LJ}\geq4098$. When $N_{LJ}=5981$. The multishell-pit NW presents. For stronger interaction ($\epsilon_{cs}$=0.035eV), the LJ atoms encapsulated by CNT form a hybrid NW for 1573$\geq N_{LJ}\geq$2845. Then it turns into multishell NWs for $N_{LJ}\geq 3471$.
For normal and stronger interactions ($\epsilon_{sc}=0.02578eV, 0.035eV$), there is a phase transition from Crystal NW (Hybrid NW) to Multishell NW, which is consistent with former study that multishell NWs should be formed with larger $N_{LJ}$.\cite{PTau} With comparison between two system with different interaction, we proved that the critical number of atoms related to the phase transition decrease with the increase of interaction.

The change of structure via  $N_{LJ}$ and  $\epsilon_{cs}$ could be understood in terms of the competition between the elastic energy of NWs and NW-CNT interface energy. On one hand, the LJ atoms tend to form fcc crystal to lower elastic/deformation energy, since any distortion on the fcc structure will result in the increase of elastic/deformation energy. On the other hand, the interaction between NW and CNT has the tendency to form the shell structure, which matches the shape of CNT well and reduce the interface energy accordingly.  The crystal NW has lower elastic energy, but higher interface energy because of the mismatch near CNT wall. The multishell NW has higher elastic energy, but lower interface energy because of well matching at NW-CNT interface. The hybrid NW is somehow in between. If $N_{LJ}$ or $\epsilon_{cs}$ is small,  the interfacial mismatch is not particularly important, the LJ atoms prefer the crystal structure.  This is the case for weaker interactions and smaller number of LJ atoms.  As $N_{LJ}$ increases, the area of the interface becomes larger and larger, correspondingly the interface energy increases if crystal NWs keep. With the accumulation of interface energy, at certain critical value of $N_{LJ}$, the structure will turn into a multishell NW, thus releasing the interface energy. Similarly, with increasing $\epsilon_{cs}$,  the interface energy increases too.

There are two particular cases, which involve the competition between the surface energy of NWs and NW-CNT interface energy.   This situation occurs at crystal-pit and multishell-pit NWs. Here NWs presents two droplet-shape pits on the side surface of NW, as compensation to disappearing surfaces of NWs by contacting two ends of nanowire to form an infinite NW.

\section{Conclusions}

A molecular dynamic simulation was performed to study the structure of LJ NWs encapsulated by CNT. The stable structure of LJ NWs confined inside CNT varies with the number of LJ atoms, the radius of CNTs, as well as the interaction between LJ NWs and CNT. For NWs confined inside (30,30) CNTs, the crystal NW, the hybrid NW, the multishell NW,  the crystal NW with two droplet-shape pits near the wall (crystal-pit NW), and  multishell NW with two droplet-shape pits near the wall (multishell-pit NW) are found. The phase boundaries among these structures are found to be affected by the interaction between LJ atoms and CNT, as well as the number of LJ atoms. For (15,15) CNTs, the multishell nanowire is the most stable structure in all cases studied here.

{\bf Acknowledgments:}  This research is supported by the Natural Science Foundation of
China, National Basic Research Program of China (973),
Shuguang and Innovation Program of Shanghai Education
Committee. The computation is performed in the Supercomputer
Center of Shanghai and ECNU.

\begin{figure}[htp]
 \centering
 \includegraphics[angle=0,width=5.5in]{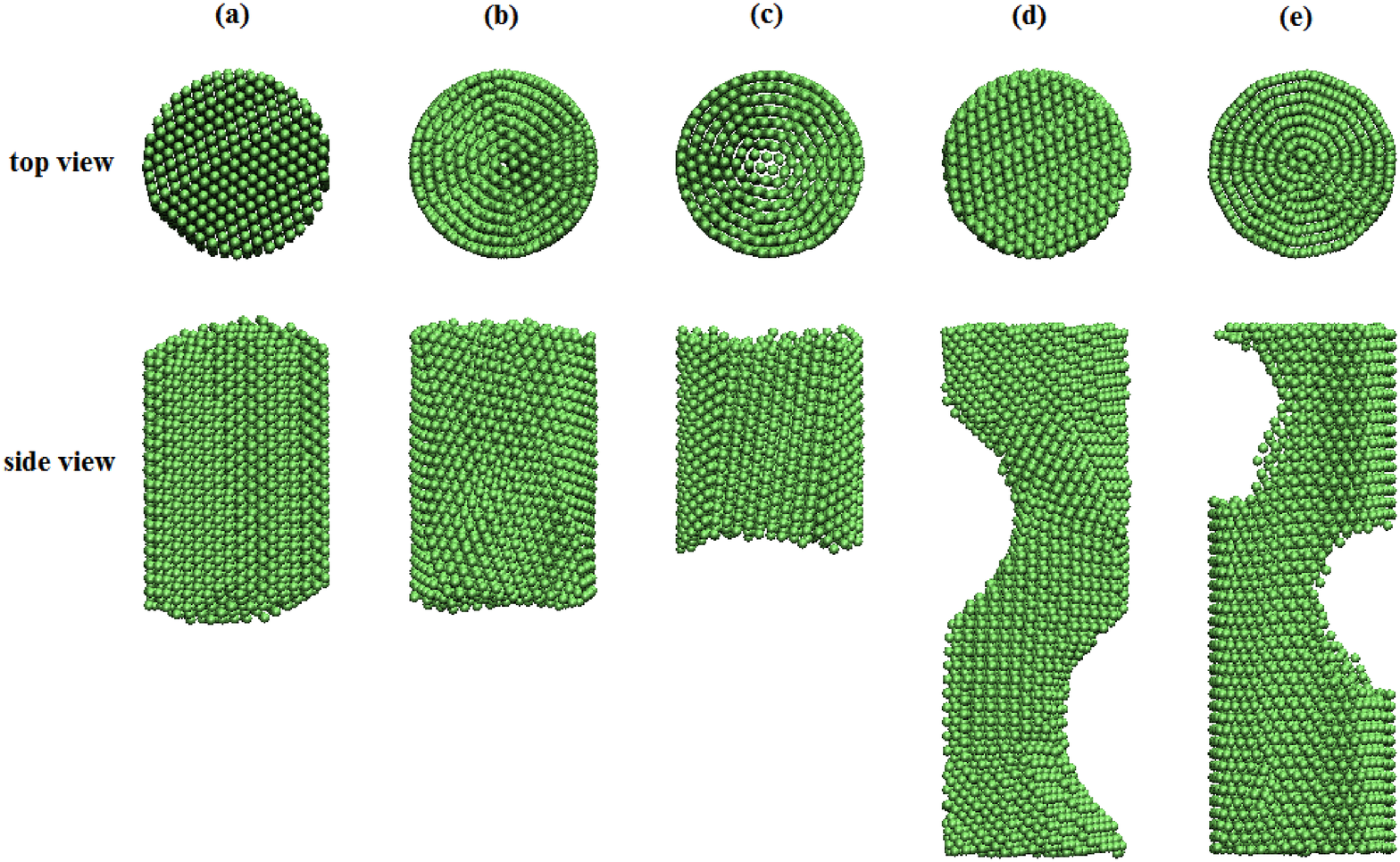}
\caption{(Color online) Top and side view of five types of NWs encapsulated by (30,30) CNTs, where carbon atoms do not show. From (a) to (e), they are the crystal NW, multishell NW, hybrid NW, crystal NW with two droplet-shape pits on the side (crystal-pit NW), and multishell NW with two droplet-shape pits on the side (multishell-pit NW), respectively.}
\end{figure}

\begin{figure}[htp]
 \centering
 \includegraphics[angle=270,width=3in]{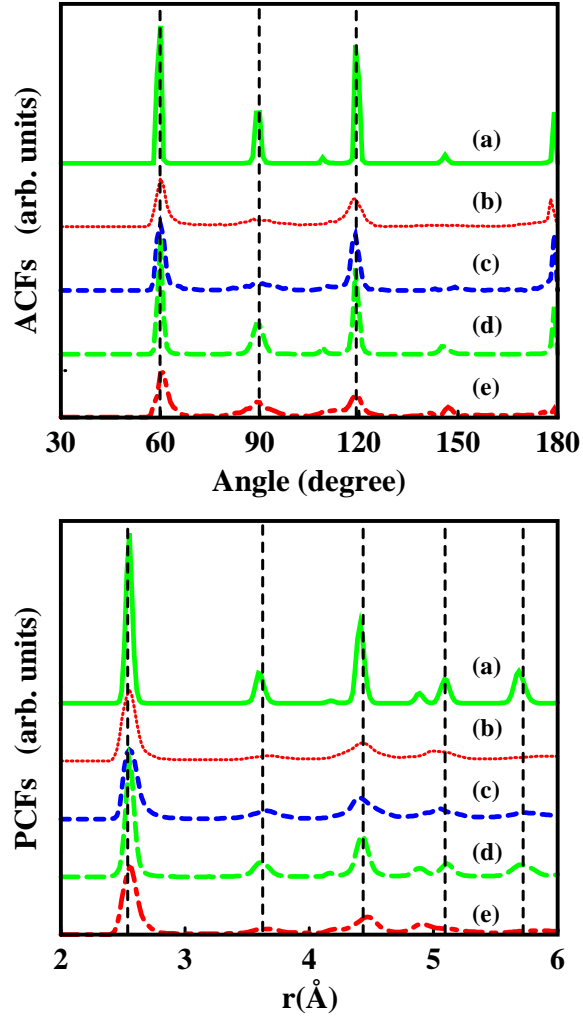}
\caption{(Color online) The angular correlation functions (ACFs) and the pair correlation functions (PCFs) of  five types of NWs encapsulated by (30,30) CNTs. The vertical lines indicate the peak position of prefect crystals.  From top to bottom, (a) green solid line: Crystal NW; (b) red dotted line: Multishell NW; (c) blue dashed line: Hybrid NW; (d) green long-dashed line: Crystal-pit NW; (f) red dashed-dotted line: Multishell-pit NW.}
\end{figure}

\begin{figure}[htp]
 \centering
 \includegraphics[angle=0,width=5in]{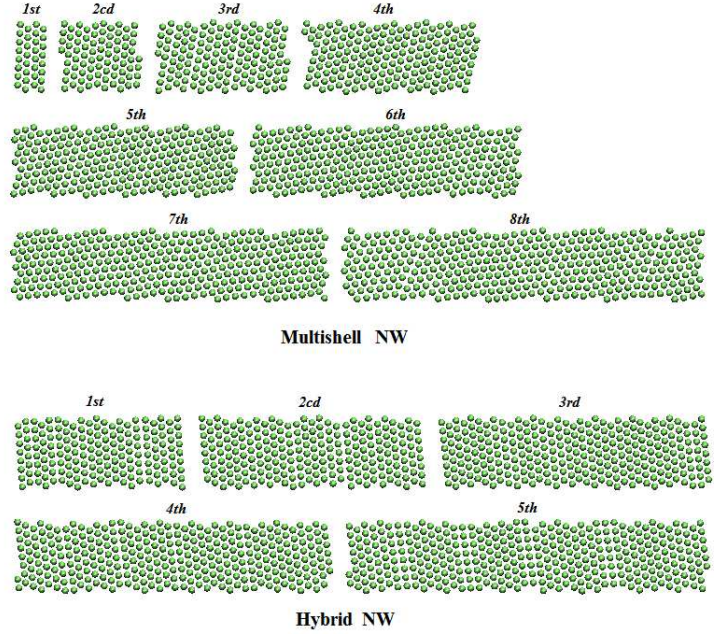}
\caption{(Color online) Upper panel: The spreading sheets of multishell NWs, where the $8th$ layer refers the outmost shell. Lower panel: The spreading sheets of  hybrid NW, where the $5th$ layer refers the outmost shell. }
\end{figure}

\begin{figure}[htp]
 \centering
 \includegraphics[angle=270,width=4in]{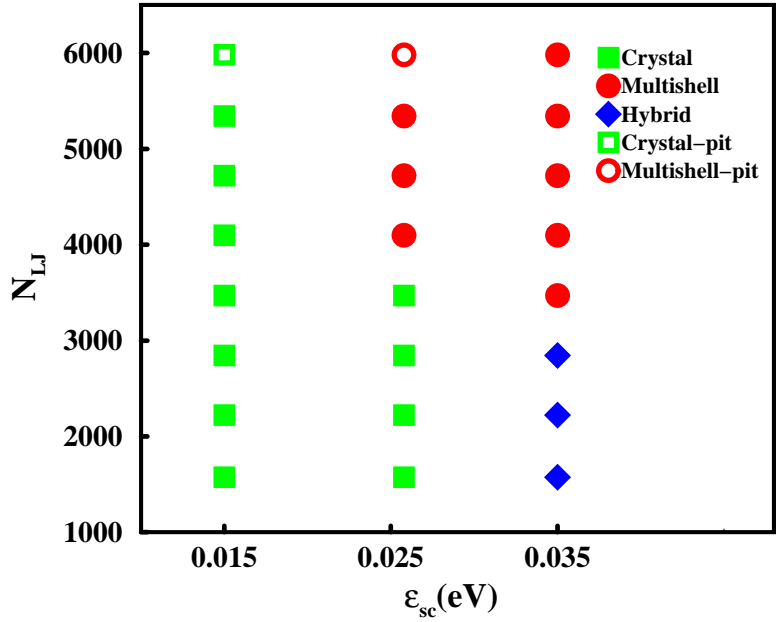}
 \caption{(Color online) The 'phase' diagram for NWs encapsulated by (30,30) CNT, namely the structure of NWs via the number of LJ atoms and the interaction between NW and CNT. }
\end{figure}

\end{document}